\documentclass{hep99}
\usepackage{epsf}

\newcommand{\GFermi}{G_{\rm F}}

\newcommand{\order}[1]{{\cal O}(#1)}
\newcommand{\bld}[1]{\boldmath{$#1$}}

\renewcommand{\Im}{{\rm Im}}

\title{Second Order QCD Corrections to the Top Decay Rate$^\dagger$
  \footnotetext[1]{Talk presented by R.H. at EPS HEP99, Tampere, Finland, July
  15-21, 1999.\\
  Work supported by DFG, Contract Ku 502/8-1, and Schweizer
  Nationalfonds.}
  \vspace{-6em}
  \begin{flushright}
    \normalsize
    \bf TTP99-41\\
    \bf BNL-HET-99/29\\
    \bf October 1999\\
    \bf hep-ph/9910339
  \end{flushright}
  \vspace{4em}
  }

\author{K.G. Chetyrkin$^1$, R. Harlander$^{1,2}$, T. Seidensticker$^1$, 
  M.~Steinhauser$^3$}
\address{$^1$University of Karlsruhe,
  D-76128 Karlsruhe, Germany\\ 
  $^2$Physics Department, Brookhaven National Laboratory, Upton, NY 11973\\
  $^3$University of Hamburg, D-22761 Hamburg, Germany}

\abstract{We report on a recent calculation of the top decay rate up to
  $\order{\alpha_s^2}$. It is based on asymptotic expansions of
  the off-shell top propagator, followed by a Pad\'e approximation in
  order to reach the physically relevant point $q^2=M_t^2$. }

\begin{document}

\maketitle
\mbox{}\thispagestyle{empty}\newpage
\mbox{}\thispagestyle{empty}\newpage
\mbox{}\thispagestyle{empty}\newpage

\title{Second Order QCD Corrections to the Top Decay Rate$^\dag$}

\maketitle
  \fntext{\dag}{
    Talk presented by R.H. at EPS HEP99.
    Work supported by DFG, Contract Ku 502/8-1, and Schweizer
    Nationalfonds.
}
\setcounter{page}{1}
\thispagestyle{empty}
\section{Introduction}
The top quark is currently the heaviest known elementary particle. Thus
top physics is a very promising field with regard to physics beyond the
Standard Model (SM). For example, its large Higgs coupling makes one
hope to learn something about the Higgs spectrum in particular, or mass
generating mechanisms in general. Furthermore, its large mass is an
important premise for decays into non-standard particles.  Another
exceptional property of the top quark is its large decay width
($\Gamma_{\rm Born} \approx 1.56$~GeV) as predicted by the SM.  Instead
of undergoing the process of hadronization, the top quark is hardly
affected by the non-perturbative regime of QCD and decays almost
exclusively into a bottom quark and a $W$ boson by weak interaction.

\section{Known results}
The $\order{\alpha_s}$ corrections~\cite{JK:twb} to the decay rate
$\Gamma(t\to bW)$ amount to $-9\%$ of the Born result. This is
comparable to the expected experimental accuracy at an NLC which is
around $10\%$.  Thus one should make sure that the QCD corrections are
reliable, in the sense that the series in $\alpha_s$ converges
sufficiently fast. This is the main motivation for investigating the
$\alpha_s^2$ corrections to this process.  The $-9\%$ from above may be
split into a contribution for $M_W=0$ ($-11\%$) and the effects induced
by a non-vanishing $W$ mass ($+2\%$).  The electroweak
corrections~\cite{twb:ew1l} at one-loop level are about $2\%$.

For the $\order{\alpha_s^2}$ corrections there exists a result for
$M_W=0$~\cite{CzaMel:twb}. It was obtained by performing an expansion
in the limit $$(M_t^2 - M_b^2)/M_t^2 \ll 1$$ and taking into account a
sufficiently large number of terms in the expansion. The result reads
\begin{equation}
\Gamma_t = \Gamma_{\rm Born}(1-0.09-0.02)\,,
\end{equation}
where the three numbers correspond to the Born, $\order{\alpha_s}$, and
$\order{\alpha_s^2}$ terms, respectively.

The aim of the calculation of~\cite{CHSS:twb} was, on the one hand, to
perform an independent check of the $\order{\alpha_s^2}$ terms at
$M_W=0$, and, on the other hand, to take into account effects induced by
a non-vanishing $W$ mass at this order.

\section{Method}\label{sec::method}
\begin{figure}
  \begin{center}
    \leavevmode
    \epsfxsize=4.cm
    \epsffile[75 275 500 460]{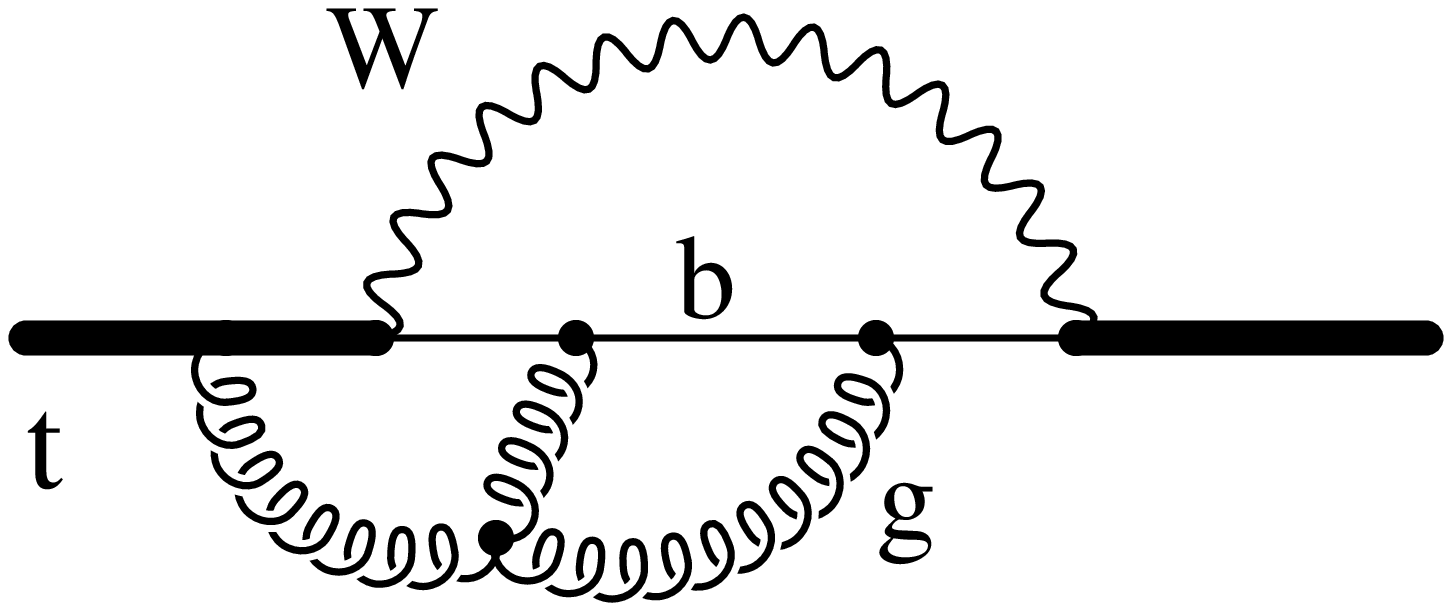}
    \hfill
    \parbox{7.cm}{
    \caption[]{\label{fig::dia}\sloppy
      Sample diagram contributing to the top quark propagator at
      $\order{\GFermi\alpha_s^2}$.
      }}
  \end{center}
\end{figure}

While in \cite{CzaMel:twb} vertex diagrams for $t\to bW$ were computed,
in \cite{CHSS:twb} the top quark decay rate was obtained via the optical
theorem which relates it to the imaginary part of the top quark
propagator $\Sigma = q\hspace{-1.15ex}/ \Sigma_{\rm V} + M_t \Sigma_{\rm S}$:
\begin{equation}
\Gamma_t \propto \Im(\Sigma_{\rm V} + \Sigma_{\rm S})\big|_{q^2=M_t^2}\,.
\end{equation}
This means that one should calculate $\Sigma$ at the point $q^2=M_t^2$
up to $\order{\GFermi\alpha_s^2}$. An example for a diagram that
contributes to this order is shown in Fig.~\ref{fig::dia}.  The $b$
quark mass can safely be set to zero, and for the moment also $M_W=0$
will be assumed.  The analytic evaluation of diagrams like the one shown
in Fig.~\ref{fig::dia} is currently neither available for general $q^2$,
nor for the special case of interest, $q^2=M_t^2$. The only limiting
cases that are accessible are $q=0$ or $M_t=0$. However, asymptotic
expansions provide an efficient tool to obtain approximate results also
away from these extreme choices.  Their application yields series in
$q^2/M_t^2$ (or $M_t^2/q^2$), with the individual coefficients
containing the non-analytic structures in terms of logarithms of
$\mu^2/q^2$ and $\mu^2/M_t^2$ ($\mu$ is the renormalization scale).
Employing the analyticity properties of the approximated function, one
obtains the region of convergence for the corresponding series. Within
this region, the full result can be approximated with arbitrary accuracy
by including sufficiently many terms in the expansion. Examples
demonstrating the quality of such approximations can be found in
\cite{CHKS:m12,HSS:zbb}.

Applying this strategy to top quark decay, in a first step one should
compute as many terms as possible in the expansion around
$q^2/M_t^2$.  Note that one cannot approach the point $q^2=M_t^2$ from
the opposite side (i.e.~$q^2/M_t^2 > 1$), because this implies top
quarks in the final state which is kinematically forbidden.  The second
step is to extrapolate the result from the small-$q^2$ region to the
point $q^2=M_t^2$. At $\order{\alpha_s}$ this could be done by
explicitly resumming the full series in $q^2/M_t^2$~\cite{CzaMel:proc}.
At $\order{\alpha_s^2}$, however, a different strategy was pursued in
\cite{CHSS:twb} by performing a Pad\'e approximation in the variables
$z=q^2/M_t^2$ and $\omega = (1-\sqrt{1-z})/(1+\sqrt{1-z})$.

Effects of a non-vanishing $W$ mass can be taken into account by applying
asymptotic expansions w.r.t.\ the relation $M_t^2\gg q^2\gg M_W^2$. In
this way one obtains a nested series in $M_W^2/M_t^2$ and
$q^2/M_t^2$. The above procedure can then be applied to each
coefficient of $M_W^2/M_t^2$ separately.

\begin{figure}
  \begin{center}
    \leavevmode
    \epsfxsize=6.cm
    \epsffile[110 265 465 540]{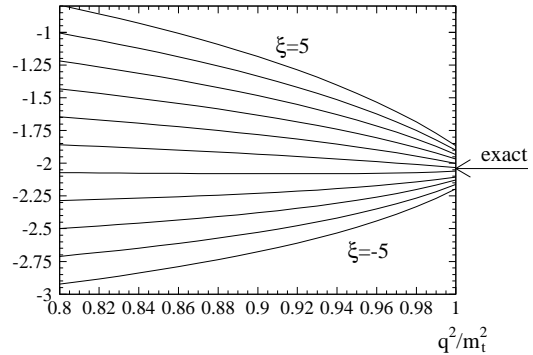}
    \hfill
    \parbox{7.cm}{
    \caption[]{\label{fig::xi}\sloppy
      $q^2$ dependence of the $[4/4]$ Pad\'e result at
      $\order{\GFermi\alpha_s}$ for different values of $\xi$.  }}
  \end{center}
\end{figure}

One of the questions one is faced with when following this approach is
gauge (in)dependence. The off-shell fermion propagator is not a gauge
invariant quantity, and only in the limit $q^2=M_t^2$ the QCD gauge
parameter $\xi$ formally drops out.  Due to the fact that one works with
a limited number of terms here, gauge parameter dependence does not
vanish exactly but is only expected to decrease gradually as soon as a
sufficiently large number of expansion terms is included. Nevertheless,
the claim is that by a reasonable choice of the gauge parameter the
predictive power of the result is preserved.  At $\order{\alpha_s^2}$
the calculation cannot be performed for general $\xi$ due to the
enormous increase in required computer resources.  Thus one cannot
decide upon the choice of $\xi$ {\it a posteriori}, but has to set it to
some definite value from the very beginning. On the other hand, it is
expected that the behavior of the $\alpha_s^2$ terms w.r.t.\ the choice
of $\xi$ is similar to the $\order{\alpha_s}$ result. Therefore, in
\cite{CHSS:twb} the gauge parameter dependence was studied in some
detail at $\order{\alpha_s}$. This study was mainly based on the
stability of the Pad\'e results $[m/n]$ upon variation of $m$ and $n$
for different values of $\xi$.  Another way to find a ``reasonable''
choice for $\xi$ is to study the $q^2$-dependence of the Pad\'e results
close to the physically relevant point $q^2=M_t^2$. The extrapolation to
$q^2=M_t^2$ is expected to work best if the variation of the
approximating function close to this point is smoothest. For some values
of $\xi$, the $q^2$ dependence of the one-loop result near $q^2/M_t^2=1$
is shown in Fig.~\ref{fig::xi}. The results of \cite{CHSS:twb} were
obtained by setting $\xi=0$.  The validity of this choice is further
justified by the perfect agreement of the approximation to the exact
result at $\order{\alpha_s}$ (see Fig.~\ref{fig::xi} and the results in
the following section).

Concerning the technical realization of the calculation it heavily
relies on automatic Feynman diagram evaluation with the help of
algebraic programs \cite{HS:review}. For details we refer
to~\cite{CHSS:twb}.

\section{Results}
The result for the top decay rate to $\order{\alpha_s^2}$ will be
written in the following way ($y\equiv M_W^2/M_t^2$, $y_0 = (80.4/175)^2$):
\begin{equation}
  \Gamma_t = \Gamma_0\left(\delta^{(0)}(y) +
    {\alpha_s\over\pi}\delta^{(1)}(y) 
    + \left({\alpha_s\over\pi}\right)^2 \delta^{(2)}(y) \right)\,,
\label{eq::gammat}
\end{equation}
with $\delta^{(0)}(y_0) = 0.885\ldots$.
Following the method described in Section~\ref{sec::method}, one obtains
$\delta^{(1)}(y_0) = -2.20(3)$ which is to be compared with the
exact number, reading $-2.220\ldots$. This demonstrates the validity and
the accuracy of the underlying approach.

At $\order{\alpha_s^2}$, the result of \cite{CHSS:twb} reads
\begin{eqnarray}
\lefteqn{\delta^{(2)}(y) = -16.7(8) + 5.4(4)\, y
+ y^2\,(11.4(5.0)} \nonumber\\&&\mbox{}+ 7.3(1)\,\ln y) + \order{y^3}
\stackrel{y=y_0}{=} -15.6(1.1).\,\,\,\,\,\,\label{eq::top2l}
\end{eqnarray}
The first number corresponds to the case of vanishing $W$ mass. It
agrees perfectly with the result of \cite{CzaMel:twb} which is
$-16.7(5)$. The uncertainty quoted in \cite{CHSS:twb} is slightly larger
than the one of \cite{CzaMel:twb} but is still negligible as compared to
the expected experimental accuracy at future colliders. The series in
$M_W^2/M_t^2$ is converging very quickly, and the total effect of the
$M_W$-suppressed terms is small. 
The errors
in (\ref{eq::top2l}) are added linearly, yielding a bigger uncertainty for
the sum than for the $M_W=0$ result. Adding the errors quadratically
instead, the uncertainty again would be $0.8$.  

\section{Estimate for \bld{b\to ul\bar\nu}}
\begin{figure}
  \begin{center}
    \leavevmode
    \begin{tabular}{c}
      \epsfxsize=5.5cm
      \epsffile[110 265 465 540]{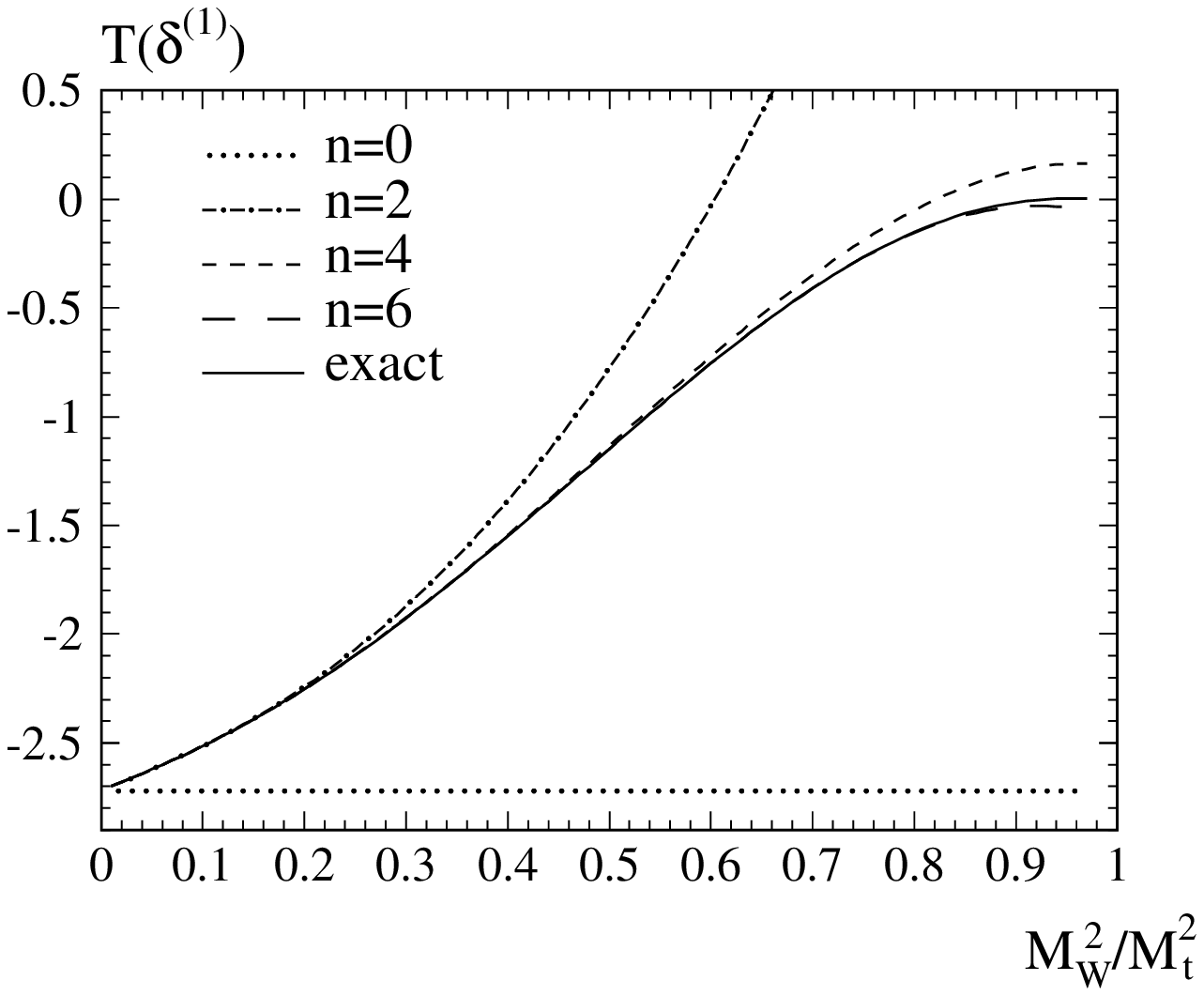}\\[1em]
      \epsfxsize=5.5cm
      \epsffile[110 265 465 540]{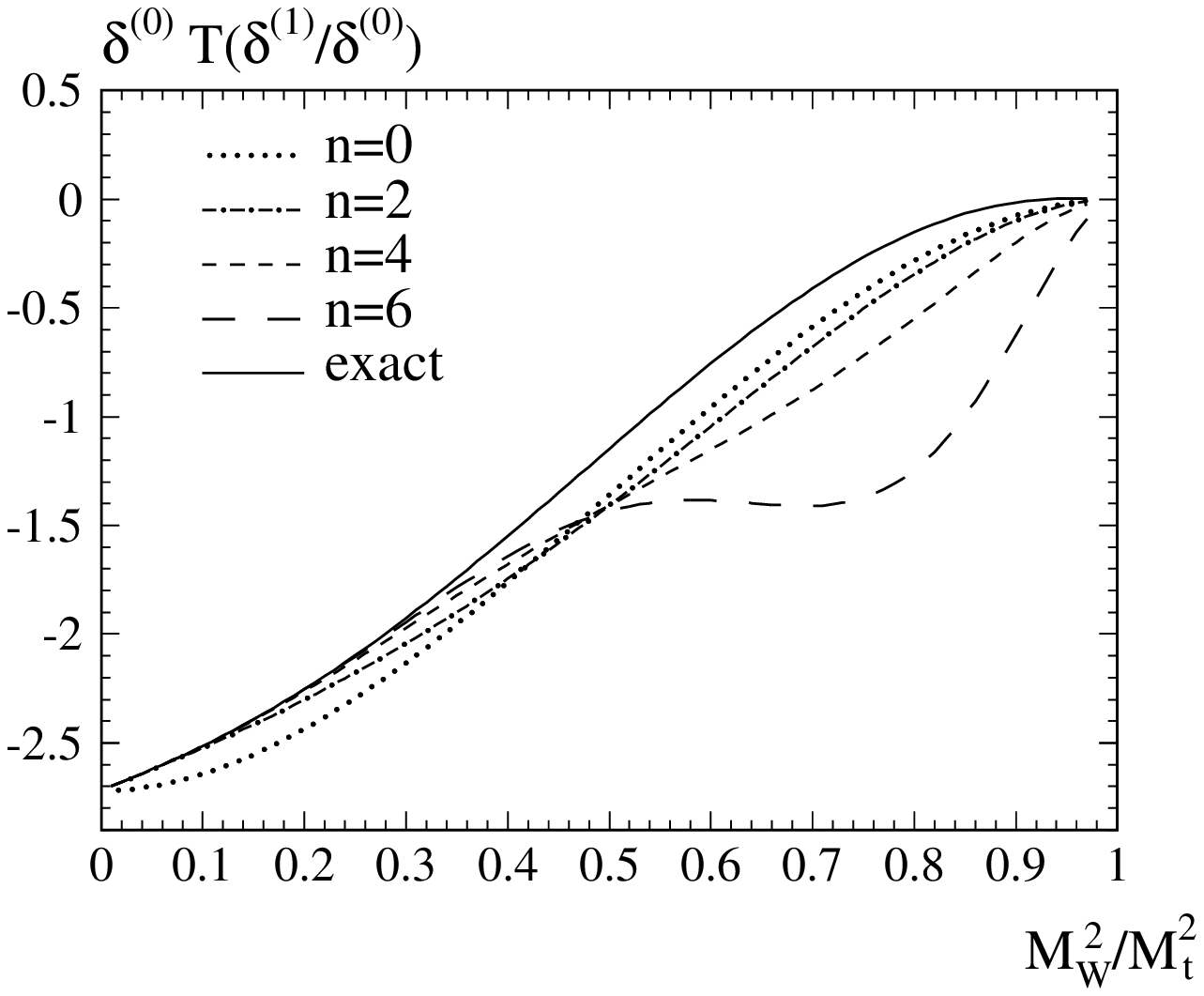}
    \end{tabular}
    \parbox{7.cm}{
      \caption[]{\label{fig::a0a1}\sloppy
        ${\rm T}(f)$ means Taylor expansion of $f$ around small
        $y=M_W^2/M_t^2$ up to order $n$ (only even orders are shown).
        The solid line represents the exact result for $\delta^{(1)}$.
        }}
  \end{center}
\end{figure}

The integration of Eq.~(\ref{eq::gammat}) over $y$ from $0$ to $1$ is
directly related to the decay rate for $b\to ul\bar \nu$ at
$\order{\alpha_s^2}$ (and with obvious modifications to the two-loop QED
corrections to $\Gamma(\mu\to e\nu\bar \nu))$. However, since the $y$
dependence of $\delta^{(2)}$ is only known up to $y^2$, it is more
promising to pull out $\delta^{(0)}(y)$ and to consider the Taylor
expansion of $\delta^{(i)}/\delta^{(0)}$ which is expected to vary only
slightly in the relevant $y$-range~\cite{CzaMel:twb}.  On the other
hand, if more terms in $y$ are included, the Taylor series of
$\delta^{(i)}/\delta^{(0)}$ becomes ill-behaved, and one should directly
expand $\delta^{(i)}$. These observations are illustrated for $i=1$ in
Fig.~\ref{fig::a0a1}. In \cite{CHSS:twb} both methods were applied, and
the results for $\Gamma(b\to ul\bar \nu)$ and $\Gamma(\mu\to e\nu\bar
\nu)$ agree with the ones of $\cite{RitStu:bmudecay}$ to about $10\%$.
This constitutes a stringent check on both calculations.

Also a direct application of the method described in
Section~\ref{sec::method} to the (four-loop!) diagrams corresponding to
$b\to ul\bar\nu$ and $\mu\to e\nu\bar \nu$ was performed in
\cite{KueSeiSte:mu}, and full agreement with $\cite{RitStu:bmudecay}$
was found.

\def\app#1#2#3{{\it Act.~Phys.~Pol.~}{\bf B #1} (#2) #3}
\def\apa#1#2#3{{\it Act.~Phys.~Austr.~}{\bf#1} (#2) #3}
\def\cmp#1#2#3{{\it Comm.~Math.~Phys.~}{\bf #1} (#2) #3}
\def\cpc#1#2#3{{\it Comp.~Phys.~Commun.~}{\bf #1} (#2) #3}
\def\epjc#1#2#3{{\it Eur.\ Phys.\ J.\ }{\bf C #1} (#2) #3}
\def\fortp#1#2#3{{\it Fortschr.~Phys.~}{\bf#1} (#2) #3}
\def\ijmpc#1#2#3{{\it Int.~J.~Mod.~Phys.~}{\bf C #1} (#2) #3}
\def\ijmpa#1#2#3{{\it Int.~J.~Mod.~Phys.~}{\bf A #1} (#2) #3}
\def\jcp#1#2#3{{\it J.~Comp.~Phys.~}{\bf #1} (#2) #3}
\def\jetp#1#2#3{{\it JETP~Lett.~}{\bf #1} (#2) #3}
\def\mpl#1#2#3{{\it Mod.~Phys.~Lett.~}{\bf A #1} (#2) #3}
\def\nima#1#2#3{{\it Nucl.~Inst.~Meth.~}{\bf A #1} (#2) #3}
\def\npb#1#2#3{{\it Nucl.~Phys.~}{\bf B #1} (#2) #3}
\def\nca#1#2#3{{\it Nuovo~Cim.~}{\bf #1A} (#2) #3}
\def\plb#1#2#3{{\it Phys.~Lett.~}{\bf B #1} (#2) #3}
\def\prc#1#2#3{{\it Phys.~Reports }{\bf #1} (#2) #3}
\def\prd#1#2#3{{\it Phys.~Rev.~}{\bf D #1} (#2) #3}
\def\pR#1#2#3{{\it Phys.~Rev.~}{\bf #1} (#2) #3}
\def\prl#1#2#3{{\it Phys.~Rev.~Lett.~}{\bf #1} (#2) #3}
\def\pr#1#2#3{{\it Phys.~Reports }{\bf #1} (#2) #3}
\def\ptp#1#2#3{{\it Prog.~Theor.~Phys.~}{\bf #1} (#2) #3}
\def\ppnp#1#2#3{{\it Prog.~Part.~Nucl.~Phys.~}{\bf #1} (#2) #3}
\def\sovnp#1#2#3{{\it Sov.~J.~Nucl.~Phys.~}{\bf #1} (#2) #3}
\def\tmf#1#2#3{{\it Teor.~Mat.~Fiz.~}{\bf #1} (#2) #3}
\def\yadfiz#1#2#3{{\it Yad.~Fiz.~}{\bf #1} (#2) #3}
\def\zpc#1#2#3{{\it Z.~Phys.~}{\bf C #1} (#2) #3}
\def\ibid#1#2#3{{ibid.~}{\bf #1} (#2) #3}

\end{document}